\documentclass[11pt]{article}

\usepackage{emp}
\usepackage{amssymb,amsmath,amsthm}
\usepackage{cite}
\usepackage{yfonts}
\usepackage[alwaysadjust]{paralist}

\newcommand{\captionfonts}{\small}
\makeatletter  
\long\def\@makecaption#1#2{%
  \vskip\abovecaptionskip
  \sbox\@tempboxa{{\captionfonts #1: #2}}%
  \ifdim \wd\@tempboxa >\hsize
    {\captionfonts \textbf{#1}\, #2\par}
  \else
    \hbox to\hsize{\hfil\box\@tempboxa\hfil}%
  \fi
  \vskip\belowcaptionskip}
\makeatother   

\newtheorem{lemma}{Lemma}
\newtheorem{example}{Example}

\newtheorem{theorem}[lemma]{Theorem}

\newtheorem{corollary}[lemma]{Corollary}

\newcommand{\C}{\mathbf{C}}
\newcommand{\F}[0]{{\mathbb{F}}}

\newcommand{\onemat}[0]{{\mathbf 1}}
\newcommand{\qbin}[2]{\left[ {#1 \atop #2} \right]}
\newcommand{\ftwo}{{\textstyle \frac{1}{\sqrt{2}}}}

\def\ket#1{|#1\rangle}

\newcommand{\bracket}[2]{\left<#1|#2\right>}

\newcommand{\nix}[1]{{}}

\newcommand{\mysection}[1]{\bigskip\noindent{\swabfamily\Large #1}\par\medskip}
\newenvironment{tale}[1]{\noindent\swabfamily\fraklines\yinitpar{#1}}{\par\ignorespacesafterend}

\newcommand{\myparagraph}[1]{\smallskip \par\noindent\textsl{#1}$\;$}

\begin{document}
\begin{empfile}

\title{\LARGE\swabfamily New Tales: of the Mean King}
\author{Andreas Klappenecker\footnotemark{}\ \ 
 and
Martin R\"otteler\footnotemark}
\date{}
\maketitle

\begin{tale}
Once upon a time on an is:land far far away there lived a mean King
who loved cats:. The King s:tarted to hate phys:icis:ts: once he
learned what had happened to Schr\"odinger's:{} cat. One evening a
terrible s:torm came on, there was: thunder and lightning, and the
rain poured down in torrents:.  Alice, a phys:icis:t, got s:tranded
during that s:torm and the King's: men captured her and brought her to
the royal laboratories:.  Alice was: told that s:he can prepare a
s:ilver atom in any s:tate of her choos:ing, then the King's: men will
s:ecretly meas:ure one of the three cartes:ian s:pin components: and
hand the atom back to her. Alice is: then free to perform any
experiment with the s:ilver atom before the King tells: her which type
of meas:urement has: been done by his: men.  Once the King reveals:
this: s:ecret, s:he mus:t immediately s:tate the correct res:ult of
the meas:urement or s:he will die a cruel death.
\end{tale}

\mysection{The Firs:t Problem}
\noindent The first problem of the King was brought to us by Aharanov,
Vaidman, and Albert~\cite{VAA:87}, although they did not
dare\footnote{This is quite understandable; the cruel deeds of the
Mean King can easily scare young physicists and we therefore do not
recommend these tales for bedtime reading. Reader discretion is
advised.}  to reveal the tale of the King.  The story was later told
by Aharanov and Englert~\cite{AE:2001a,AE:2001b} and we retold the
tale.  In more modern terms, Alice has the problem to prepare a
quantum bit in a certain state, then the King's men perform a von
Neumann measurement with respect to one of the Pauli spin observables
$\sigma_x$, $\sigma_y$, or $\sigma_z$. Alice can then perform another
measurement.  Once the King reveals which observable was measured, she
has to quickly find the answer.
\smallskip

\begin{tale}
Alice thought about the problem. She quickly realized that it is: too
ris:ky to prepare the s:tate of a s:ingle s:ilver atom, but then s:he
had an idea.  She as:ked Tweedledee and Tweedledum to help her prepare
a pair of s:ilver atoms: in a Schr\"odinger cat s:tate.  After they
prepared the s:tate, the King's: men went to meas:ure one atom. Once
the atom was: returned, Alice meas:ured the s:ys:tem. To the great
annoyance of the King, s:he was: able to correctly gues:s: the King's:
obs:ervation. Alice was: s:et free and tried to live happily ever
after.
\end{tale}
\smallskip

We do not exactly know how Alice solved the problem, but we give a
possible explanation. Recall that the spin matrices 
$\sigma_z$, $\sigma_x$, and $\sigma_y$ are complementary observables; hence, 
their eigenbases 
$$ 
B_0 = \{ v_{0,1}, v_{0,2}\}, \quad 
B_1 = \{ v_{1,1}, v_{1,2}\}, \quad 
B_2 = \{ v_{2,1}, v_{2,2}\},
$$ 
are mutually unbiased orthonormal bases of $\C^2$.  The particular nature of
these bases is not really relevant, but the property that $|\langle
v_{a,b}| v_{a',b'}\rangle|^2=1/2$ holds when $a\neq a'$ is crucial for
our argument. One consequence is that 
$$ \varphi = \textstyle 
\ftwo v_{a,1}\otimes\overline{v_{a,1}} + 
\ftwo v_{a,2}\otimes\overline{v_{a,2}} \in \C^2\otimes \C^2 
$$ is the same state for all $a\in \{0,1,2\}$, which is in fact 
maximally entangled. Alice prepares the two silver atoms in this
state. If the King's men perform a von Neumann measurement with
respect to, say, the basis $B_a$ and observe the value $b$, then the
state of the two silver atoms collapses to $v_{a,b}\otimes
\overline{v_{a,b}}$.

The trick is that Alice can set up a measurement such that she will
learn a function $f\colon \{0,1,2\}\rightarrow \{1,2\}$ that correctly 
maps the selected basis to the observed values. 
For instance,  Alice can perform a von Neumann measurement 
with respect to the basis 
$$ \psi_{f_k} = -\varphi + \ftwo \sum_{a=0}^2 v_{a,f_k(a)}\otimes
\overline{v_{a,f_k(a)}},\qquad k\in \{1,2,3,4\},$$ where the
functions $f_k\colon \{0,1,2\}\rightarrow \{1,2\}$ are given by
$$ 
\begin{array}{l@{,\qquad}l@{,\qquad}l}
f_1(0)= 1 & f_1(1)= 1 & f_1(2)= 1, \\
f_2(0)= 1 & f_2(1)= 2 & f_2(2)= 2, \\
f_3(0)= 2 & f_3(1)= 1 & f_3(2)= 2, \\
f_4(0)= 2 & f_4(1)= 2 & f_4(2)= 1. 
\end{array}
$$ 
For example, suppose that the King's men have measured in the basis
$B_a$, $a=2$, and observed the value $b=1$. If Alice measures the
resulting state in the basis $\{\psi_{f_1}, \psi_{f_2}, \psi_{f_3}, 
\psi_{f_3}\}$, then she will learn the function
$$ 
\begin{array}{l@{\quad\text{with probability}\quad}l}
f_1 & |\langle v_{2,1}\otimes \overline{v_{2,1}}\,|\psi_{f_1}\rangle|^2 =1/2,\\
f_2 & |\langle v_{2,1}\otimes \overline{v_{2,1}}\,|\psi_{f_2}\rangle|^2 =0,\\
f_3 & |\langle v_{2,1}\otimes \overline{v_{2,1}}\,|\psi_{f_3}\rangle|^2 =0,\\
f_4 & |\langle v_{2,1}\otimes \overline{v_{2,1}}\,|\psi_{f_4}\rangle|^2 =1/2.
\end{array}
$$ 
If the King finally reveals to her that his men have chosen the basis
$a=2$, then she will predict, either way, the value
$f_1(2)=f_4(2)=1$. In fact, it can be checked that this von Neumann
measurement allows Alice to always correctly predict the observation
made by the King's men, regardless of the choice of $a$ and $b$! We
will reveal the reason for this most curious behavior in the next
section, but we encourage the reader to check our claim.
\smallskip

\textit{Remark.} There exists a variation of the first problem by
Hayashi, Horibe, and Hashimoto~\cite{HHH:04} that does not require
mutually unbiased bases. Another variation can be obtained by the
method described in the section on the King's third problem.

\mysection{The Second Problem}
\begin{tale}
She tried to leave the kingdom as: s:oon as: pos:s:ible. When
Alice finally reached the s:hore, s:ome of the King's: guards:
captured her, again! `What do you want?', s:aid Alice. One guard
replied: `We take you into cus:tody, becaus:e you are s:till a
phys:icis:t'. Alice finally unders:tood why the monarch was: known as:
the `Mean King'. The King explained to Alice that the firs:t problem was: too
eas:y. Then the King s:aid `I will hand you an atom with a prime power q of
different bas:is: s:tates:. You can prepare it in any s:tate, then my
men will s:ecretly meas:ure the atom with one of q+1 complementary
obs:ervables:'.  He was: leaning forward and s:aid `You can perform
one more experiment, but then you have to gues:s: the obs:ervation
made by my men, or...'. In s:pite of dark foreboding, Alice replied
s:ternly `Fine!'. 
\end{tale}
\smallskip

The second problem was developed in a series of papers. Aharonov and
Englert discussed a solution if the atom has a prime number of levels,
see~\cite{AE:2001a} and \cite{AE:2001b}. Aravind was the first to find
a solution of the Mean King problem for atoms with a prime power of
basis states~\cite{Aravind:2003}, followed by Durt~\cite{Durt:2004}
and very recently by Hayashi, Horibe, and Hashimoto~\cite{HHH:2005}.
The latter approach is based on maximal sets of mutually orthogonal
latin squares and turns out to be equivalent to our exposition,
although it was developed independently. We use a more geometric approach
based on affine planes.
\medskip

\begin{tale} For s:everal days: Alice was: frantically s:cribbling 
and drawing on parchment. She certainly did not want to make any
mis:take. Finally, s:he had found a geometric s:olution. She as:ked
Tweedledee and Tweedledum to help her in the royal laboratories:. They
prepared the s:tate, the King's: men meas:ured and returned the atom.
Alice then performed another experiment and after the King revealed the
meas:urement bas:is:, s:he correctly gues:s:ed the value that has:
been obs:erved by the King's: men.
\end{tale}
\smallskip

\myparagraph{Designs.}  We recall some terminology from combinatorial
design theory, see~\cite{BJL:99I,vanLindt:01,Stinson:2003} for
details.  Let $X$ be a finite set of $v$ points. A $(v,k,\lambda)$
design over $X$ is a family $\mathcal{B}$ of $k$-element subsets of
$X$, called blocks, such that every pair of distinct points is
contained in exactly $\lambda$ blocks. A consequence of the definition
is that a point lies in $r=(v-1)\lambda/(k-1)$ blocks and there is a
total of $b=vr/k$ blocks. Sometimes we list all five parameters and
talk about a $(v,b,r,k,\lambda)$ design to save the reader the trouble
to derive the parameters $b$ and $r$.

A parallel class of a $(v,k,\lambda)$ design over $X$ is a subset of
disjoint blocks whose union is $X$. The design $(X,\mathcal{B})$ is
called resolvable if there exists a partition of $\mathcal{B}$ into
parallel classes. A resolvable $(v,b,r,k,\lambda)$ design consists of
$v/k$ blocks, so $v\equiv 0\bmod k$; it has $r$ parallel classes,
because a point occurs in $r$ blocks and each class contains a point
exactly once. The reader familiar with \cite{GHW:2003} might notice
that parallel classes and ``striations'' actually are the same
objects.
\smallskip

\textit{Remark.} Some authors refer to our $(v,k,\lambda)$ designs as
simple $(v,k,\lambda)$ balanced incomplete block designs, but we
prefer brevity, following~\cite{jukna:01}.

\myparagraph{Affine Planes.}  For the second problem, we need a more
general set of functions in our construction of Alice's
measurement. We will obtain this set of functions from an affine plane
of order $n$.

An $(n^2,n^2+n,n+1,n,1)$ design is called an \textsl{affine plane} of
order $n$. For affine planes, one usually uses a more geometric
language and refers to blocks as lines. In other words, an affine
plane of order $n$ has $n^2$ points, $n^2+n$ lines, and each line
contains $n$ points.

An affine plane of order $n$ is the prototype of a resolvable design.
It is possible to partition the set of lines into $n+1$ parallel
classes that contain $n$ disjoint (parallel) lines each. A small
example might help to convey the main idea. 

\begin{example}\label{example}
The affine plane of order $2$ consists of a set of $4$ points and a
family of 6 lines. The figure below illustrates this affine plane: 
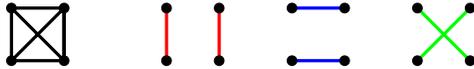
\begin{center}
\begin{emp}(50,50)
  def drawfill text t = 
    fill t; 
    draw t; 
  enddef;
  path p; 
  p := (0,0)--(0,1)--(1,1)--(1,0)--(0,0)--(1,1)--(1,0)--(0,1); 

  pickup pencircle scaled 1.25; 
  draw p scaled 7mm;
  drawfill fullcircle scaled 1mm shifted (0,0); 
  drawfill fullcircle scaled 1mm shifted (0,7mm);
  drawfill fullcircle scaled 1mm shifted (7mm,0);
  drawfill fullcircle scaled 1mm shifted (7mm,7mm);
\end{emp}
\quad\qquad
\begin{emp}(50,50)
  pickup pencircle scaled 1.25; 
  draw (0,0)--(0,7mm) withcolor red;
  draw(7mm,0)--(7mm,7mm) withcolor red;
  drawfill fullcircle scaled 1mm shifted (0,0); 
  drawfill fullcircle scaled 1mm shifted (0,7mm);
  drawfill fullcircle scaled 1mm shifted (7mm,0);
  drawfill fullcircle scaled 1mm shifted (7mm,7mm);
\end{emp}
\qquad
\begin{emp}(50,50)
  pickup pencircle scaled 1.25; 
  draw (0,0)--(7mm,0) withcolor blue;
  draw(0,7mm)--(7mm,7mm) withcolor blue;
  drawfill fullcircle scaled 1mm shifted (0,0); 
  drawfill fullcircle scaled 1mm shifted (0,7mm);
  drawfill fullcircle scaled 1mm shifted (7mm,0);
  drawfill fullcircle scaled 1mm shifted (7mm,7mm);
\end{emp}
\qquad
\begin{emp}(50,50)
  pickup pencircle scaled 1.25; 
  draw (7mm,0)--(0,7mm) withcolor green;
  draw(0,0)--(7mm,7mm)  withcolor green;
  drawfill fullcircle scaled 1mm shifted (0,0); 
  drawfill fullcircle scaled 1mm shifted (0,7mm);
  drawfill fullcircle scaled 1mm shifted (7mm,0);
  drawfill fullcircle scaled 1mm shifted (7mm,7mm);
\end{emp}
\end{center}
The three parallel classes of this affine plane are depicted on
the right in the colors red, blue, and green. Put differently, it is a $(4,2,1)$ design $(X,\mathcal{B})$ with four points $X=\{(1,1),(1,2),(2,1),(2,2)\}$ and six lines 
$\mathcal{B}= L_0\cup L_1\cup L_2$ given as a union of the three parallel classes
$$\begin{array}{lcl}
L_0 &=& \{\{ (1,1),(1,2)\}, \{(2,1),(2,2)\}\},\\ 
L_1 &=& \{\{(1,1),(2,1)\}, \{ (1,2),(2,2)\}\}, \\
L_2 &=& \{\{(1,1),(2,2)\}, \{ (1,2),(2,1)\}\}. 
  \end{array}
$$ 
\end{example}

\myparagraph{Functions.} We will now derive a set of $n^2$ functions
from an affine plane of order $n$. As a guiding example, we will see
how the four functions $f_1,\dots, f_4$ from the first section can be
derived from Example~\ref{example}.

Let $X=\{ (x,y)\,|\, 1\le x,y\le n\}$ be a set of $n^2$ points.
Suppose that $(X,\mathcal{B})$ is an affine plane of order $n$. 
We denote by $L_0,\dots,L_n$ the $n+1$ parallel classes of lines that
partition the set of lines, $\mathcal{B}= L_0\cup L_1\cup\cdots  \cup L_n$. 

Without loss of generality, we assume that the lines in the parallel
class $L_0$ are given by
$$\ell_x=\{(x,y)\,|\, 1\le y\le n\}, \quad\text{ where }\quad x\in
\{1,\dots,n\}.$$ Indeed, we can always achieve this by renaming the
points in $X$. 

Recall that two lines from different parallel classes meet in one
point. In particular, if we choose a line $\ell$ that is not contained
in $L_0$, then $\ell \cap \ell_x\neq \emptyset$ for all $1\le x\le
n$. Therefore, for each $x$ in the range $1\le x\le n$ there exists an
integer $y_x$ such that $(x,y_x)\in \ell$.

Given a parallel class $L_i$, with $1\le i\le n$, and a line $\ell$ in
$L_i$, we define a function $f_{i,\ell}\colon \{0,1,\dots,n\}
\rightarrow \{1,\dots,n\}$ by 
$$ f_{i,\ell}(a) = \left\{
\begin{array}{ll}
i & \text{if } a=0\\
b & \text{if } (a,b) \in \ell, a \neq 0
\end{array}
\right.
$$ 
Our convention for labeling the points in $X$ ensures that this
function is defined for all $a\in \{0,\dots, n\}$.

We denote by $\Delta_{f,g}$ the collision set of two functions $f$ and
$g$, defined by
$$\Delta_{f,g}= \{ x\mid f(x)=g(x)\}.$$

\begin{lemma}\label{affine}
If $f,g \in \{f_{i,\ell} \mid 1\le i\le n, \ell \in L_i\}$ are distinct functions, then the functions have exactly one collision, $|\Delta_{f,g}|=1$. 
\end{lemma}
\begin{proof}
Suppose that $f=f_{i,\ell}$ and $g=f_{i^*,\ell^*}$.  If $i=i^*$, then
$\ell$ and $\ell^*$ are parallel lines; hence, $f(a)\neq g(a)$ for
$1\le a\le n$, and a single collision is given by $f(0)=g(0)=i$.

If $i\neq i^*$, then $f(0)\neq g(0)$. Furthermore, $\ell$ and $\ell'$
are lines from distinct parallel classes, so $\ell$ and $\ell'$ have
exactly one point in common; hence, $f$ and $g$ have once again one collision. 
\end{proof}

\setcounter{example}{0}
\begin{example}[cont'd] 
In the affine plane of order 2, we choose the parallel
class $L_1= \{ \ell_1, \ell_2\}$ that consists of the lines 
$\ell_1 =\{(1,1),(2,1)\}$ and $\ell_2= \{ (1,2),(2,2)\}.$ 
The associated functions $f_1 = f_{1,\ell_1}$ and 
$f_2 = f_{1,\ell_2}$ are given by 
$$ \begin{array}{l@{\quad}l@{\quad}l}
f_{1,\ell_1}(0)=1, & 
f_{1,\ell_1}(1)=1, & 
f_{1,\ell_1}(2)=1, \\
f_{1,\ell_2}(0)=1, & 
f_{1,\ell_2}(1)=2, & 
f_{1,\ell_2}(2)=2. 
   \end{array}
$$ 
\end{example}

\myparagraph{Reconstruction.}  We will now derive an orthonormal basis
of $\C^n\otimes \C^n$. The basis is determined by the $n^2$ functions
that we have obtained in Lemma~\ref{affine} from an affine plane. It
turns out that this basis allows Alice to extract all necessary
information so that she can guess the value that has been observed by
the King's men with certainty.

Let $B_a=\{v_{a,b}\,|\, 0\le b<n\}$, with $0\le a\le n$, denote a set
of $n+1$ mutually unbiased bases \cite{Ivanovic:81,WF:89} of $\C^n$;
that is, each set $B_a$ is an orthonormal basis of $\C^n$ and the
squared-modulus of the inner product between vectors of different
bases satisfies $|\langle v_{a,b}|v_{a',b'}\rangle|^2=1/n$ for $a\neq
a'$ and all $b$ and $b'$.  Given a function $f\colon
\{0,\ldots,n\}\rightarrow \{1,\ldots,n\}$, we define the vectors $$
\varphi = \frac{1}{\sqrt{n}}
\sum_{b=1}^n v_{0,b}\otimes \overline{v_{0,b}}\,,\qquad
\gamma_f = \frac{1}{\sqrt{n}}
\sum_{a=0}^n v_{a,f(a)}\otimes \overline{v_{a,f(a)}}, $$ 
and $\psi_f = - \varphi + \gamma_f$. 

\begin{lemma}\label{onb}
Let $f, g : \{0,1,\ldots,n\} \rightarrow \{1,\ldots,n\}$ be functions and 
denote by $\psi_f$ and $\psi_g$ the associated vectors in
$\C^n\otimes \C^n$. The inner product $\langle \psi_f | \psi_g
\rangle = 0$ if and only if\/ $|\Delta_{f,g}|=1$. Furthermore, $\bracket{\psi_f}{\psi_f}=1$. 
\end{lemma}
\begin{proof}
If $a\neq a^\prime$ then $\langle v_{a,b}\otimes \overline{v_{a,b}} |
v_{a^\prime,b^\prime}\otimes \overline{v_{a^\prime,b^\prime}}\rangle=1/n$. 
Furthermore, we note that 
$\bracket{v_{a,b}\otimes \overline{v_{a,b}}}{\varphi}=1/\sqrt{n}$ 
for $0\le a\le n$ and $1\le b\le n$. 
We obtain 
\begin{eqnarray}
\bracket{\psi_f}{\psi_g} & = &
\bracket{\varphi}{\varphi} - \bracket{\varphi}{\gamma_g}
- \bracket{\gamma_f}{\varphi} + \bracket{\gamma_f}{\gamma_g}
\nonumber \\
& = & 1 - 2 \frac{(n+1)}{n} +  \frac{1}{n} \sum_{a=0}^n
\sum_{a^\prime=0}^n |\bracket{v_{a,f(a)}}{v_{a^\prime,g(a^\prime)}}|^2
\nonumber \\
& = & 1 - 2 \frac{(n+1)}{n} + \frac{1}{n} \bigg( |\Delta_{f,g}| \cdot 1
+ n(n+1) \cdot \frac{1}{n} \bigg) \nonumber \\
& =& \frac{n-2(n+1)+|\Delta_{f,g}|+n+1}{n} = \frac{|\Delta_{f,g}|-1}{n}.\label{l
ast}
\end{eqnarray}
Hence, $\bracket{\psi_f}{\psi_g} = 0$ if and only if
$|\Delta_{f,g}|=1$. Furthermore, from this we also obtain that
$\bracket{\psi_f}{\psi_f} = 1$.
\end{proof}

\begin{corollary}\label{basis} 
If a complete set of $n+1$ mutually unbiased bases exists in $\C^n$
and an affine plane of order $n$ exists, then there exist a set $S$ of $n^2$
functions $\{0,1,\dots,n\}\rightarrow \{1,\dots,n\}$ such that 
$$ \{ \psi_f\,|\, f\in S\}$$
forms an orthonormal basis of $\C^n\otimes \C^n$. 
\end{corollary}
\begin{proof}
If an affine plane exists, then Lemma~\ref{affine} shows that $n^2$
functions that have exactly one collision in common.  It follows from
Lemma~\ref{onb} that the corresponding states form an orthonormal
basis of $\C^n\otimes \C^n$.
\end{proof}

The crucial property of a von Neumann measurement with respect to the basis
$\{\psi_f\,|\, f\in S\}$ is that one can extract information from the
state returned by the King's men thanks to the following lemma: 
\begin{lemma}\label{extraction}
$\langle v_{a,b}\otimes \overline{v_{a,b}}| \psi_f\rangle\neq 0$ if
and only if $f(a)=b$.
\end{lemma}
\begin{proof}
By definition of $\psi_f$, we have
$$
\begin{array}{lcl}
\langle v_{a,b}\otimes \overline{v_{a,b}}| \psi_f\rangle &=& \displaystyle
-\bracket{v_{a,b}\otimes \overline{v_{a,b}}}{\varphi}
+\frac{1}{\sqrt{n}}\sum_{a'=0}^n |\langle v_{a,b}|v_{a',f(a')}|^2\\
&=& \displaystyle -\frac{1}{\sqrt{n}} + \frac{n}{\sqrt{n}}\times \frac{1}{n} 
+ \frac{1}{\sqrt{n}}[f(a)=b],
\end{array}
$$
where the last expression is the Iverson-Knuth bracket that is 1 if $f(a)=b$ 
and 0 otherwise. 
\end{proof}
\smallskip

\textit{Remark.} If $n$ is a power of a prime, then there exists both
an affine plane of order $n$ and a maximal set of $n+1$ mutually
unbiased bases; see, for instance,~\cite{BJL:99I,Stinson:2003} and
\cite{Ivanovic:81,WF:89}. If $n$ is not a power of a prime, then
neither an affine plane of order $n$ nor $n+1$ mutually unbiased bases
are known to exists. There exist some speculations that a set of $n+1$
mutually unbiased bases exist in $\C^n$ if and only if an affine plane
of order $n$ exists, but there is little evidence in support of such a
claim and it is doubtful whether one should elevate this to a
conjecture.

\nix{
\textit{Remark.}  In the very recent preprint~\cite{HHH:2005} it was
claimed that the Mean King problem can be solved if and only if $n$ is
a power of a prime; unfortunately, the argument is flawed and there
does not seem to exist an easy way to fix the proof.
}

\myparagraph{Summary.} 
Let $B_a=\{v_{a,b}\,|\, 0\le b<n\}$, with $0\le a\le n$, denote a set
of $n+1$ mutually unbiased bases. Suppose that the King's men will
perform a von Neumann measurement with respect to one of the $n+1$
bases. Then Alice can guess the outcome of the measurement by the
following procedure:

\begin{itemize}
\item
Alice starts by preparing the state
\[
\varphi = \frac{1}{\sqrt{n}} \sum_{b=1}^{n} v_{a,b} \otimes 
\overline{v_{a,b}}.
\]
Note that we obtain the same state for each $a$ in $0\le a\le n$. 
\item
The King's men perform a measurement with respect to one basis $B_a$.
Suppose that the result is $b$, then the state collapses to
$v_{a,b}\otimes \overline{v_{a,b}}$. 
The King's men hand the atom back to Alice but do not
yet reveal the measurement basis nor the outcome of the measurement. 

\item 
Alice performs a von Neumann measurement on $\C^n \otimes \C^n$ with
respect to the basis $\{\psi_f\,|\, f\in S\}$ given in
Corollary~\ref{basis}.  By Lemma~\ref{extraction}, we have $\langle
v_{a,b}\otimes \overline{v_{a,b}} | \psi_f \rangle \neq 0$ if and only if
$f(a) = b$. Put differently, the outcome of Alice's measurement is a
function $f$ such that $f(a) = b$.

\item
Finally, the King's men reveal the label $a$ of the basis $B_a$. 
\item
Alice can respond with the value $b$ such that $f(a)=b$, and 
the outcome $b$ is exactly what the King's men have observed. 
\end{itemize}

\begin{theorem}
If an affine plane of order $n$ and a set of $n+1$ mutually unbiased
bases in $\C^n$ exist, then Alice can solve the second problem of the
King with certainty.
\end{theorem}

\mysection{The Third Problem} 
\begin{tale} Having s:ucces:s:fully s:olved the King's: problems:, 
Alice as:s:umed that s:he might now leave the is:land and return home
to s:afety. The King on the other hand was: frus:trated s:ince his:
challenges: have been s:olved s:o eas:ily. He lamented for s:everal
days: and finally told his: woes: to his: wife. What s:he conveyed to
him came as: a big s:urpris:e: Envious: of the gifted phys:icis:t who
s:pends: s:o much quality time with her hus:band, s:he had s:ent a
s:py to Alice's: chamber to obs:erve her experiments:. The s:py
reported that s:ome part of Alice's: s:tate actually remained in
Alice's: lab and was: later us:ed for the recons:truction of the
King's: meas:urement. 
\smallskip

Furious: with anger about this: trickery, the King arres:ted Alice
once again and gave her a new challenge. She has: to prepare a quantum
s:tate of her own liking involving three s:ilver atoms:. They have to
be handed over to the King and s:he would not be allowed to have any
quantum s:tate left in her pos:s:es:s:ion which could be entangled
with the three atoms:. Then the King s:ecretly picks: an arbitrary
one of thes:e atoms: and meas:ures: it either one of three
complementary bas:es:. Subs:equently, the s:tate is: returned to
Alice, who is: allowed to perform one experiment on it. Afterwards:,
the King reveals: which bit has: been meas:ured and Alice immediately
has: to ans:wer with the correct outcome or els:e s:he will die an
even more cruel death.
\end{tale}
\smallskip

We now turn to a generalization of the Mean King's problem where the
vectors do not necessarily come in groups of mutually unbiased bases
and in which we consider more general combinatorial designs than
affine planes. 

\myparagraph{Affine Resolvable Designs.} 
For a general design no restriction is made about the number of points
in which two blocks can intersect. A particularly interesting case
arises when any two blocks either intersect in the same number of
points or do not intersect at all.

A $(v,k,\lambda)$ design is called an \textsl{affine resolvable
design}, or simply an \textsl{affine design}, if it is a resolvable
design and any two nonparallel blocks intersect at a fixed number 
$m$ of points. 

\newpage
\begin{lemma}\label{parameters}
Let $(X,\mathcal{B})$ denote an affine $(v,b,r,k,\lambda)$ design such
that nonparallel blocks intersect at $m$ points. Then the parameters satisfy the relations: \rm
\begin{compactenum}[(i)]
\item $m=k^2/v$;
\item $\lambda(v-k)=k(k-1)$;
\item $r=k+\lambda$;
\item $b=v+r-1$.
\end{compactenum}
\end{lemma}
\begin{proof}
The relations are standard facts about the parameters of affine designs,
see~\cite{BJL:99I,Stinson:2003,Wallis:88}.
\end{proof}
\smallskip

In the Euclidean plane there is exactly one line which passes through
a given point and is parallel with a fixed line; we need the
combinatorial analogue of this geometric fact. 
\begin{lemma}\label{blockPoint} 
Let $(X, \mathcal{B})$ be an affine resolvable design, $B\in
\mathcal{B}$ and $p\in X$ a point such that $p \notin B$. There
exists precisely one block $C\in \mathcal{B}$ such that $B \cap C =
\emptyset$ and $p \in C$.
\end{lemma}
\begin{proof}
Since $X$ is resolvable, we have a resolution of the blocks in
$\mathcal{B}$ into parallel classes. The block $B$ is contained in one
parallel class, say in~$\mathcal{C}$. Each parallel class partitions~$X$;
hence, there is a block $C_0 \in {\cal C}$ such that $p \in C_0$.
Since $p\notin B$, it follows that $B \cap C_0 = \emptyset$.

Seeking a contradiction, we suppose that there is another block $C_1
\in {\cal B}$ with $C_1 \not= C_0$ such that $p \in C_1$ and $B \cap
C_1 = \emptyset$. Since $C_0$ and $C_1$ are both parallel to $B$ they
have to be in the same parallel class; hence, $C_0 \cap C_1 =
\emptyset$ in contradiction to $p \in C_0 \cap C_1$.
\end{proof}

\myparagraph{Examples.} 
Basically two constructions of affine resolvable designs are known:
designs coming from affine geometries and Hadamard designs. The
following table summarizes the properties of these two classes:
\begin{table}[h]
\centerline{
\begin{tabular}{|c||c|c|c|c|c|}
\hline
Design & $v$ & $b$ & $r$ & $k$ & $\lambda$ \\ \hline\hline 
Affine plane & $q^2$ & $q^2 + q$ & $q+1$ & $q$ & $1$ \\
Affine space & $q^m$ & $q^{m-d} \qbin{m}{d}_q$ & $\qbin{m}{d}_q$ & 
$q^d$ & $\qbin{m-1}{d-1}_q$ \\
Hadamard designs & $4m$ & $8m-2$ & $4m-1$ & $2m$ & $2m-1$ \\
\hline
\end{tabular}}
\caption{Parameters of affine designs. In this table $m\geq 1$ and $1 \leq d < m$. The Gaussian $q$-binomial
coefficient \hbox{$\qbin{m}{d}_q$} equals the number of
$d$-dimensional subspaces of $\F_q^m$,
see~\cite{kac:02}. Constructions for affine planes and spaces with the
parameters given in this table are known when $q$ is a power of a
prime; it is conjectured that Hadamard designs exist for all $m\geq
1$.}
\end{table}

\myparagraph{Representation.}  In the following, we will relate the
discrete combinatorial structures to vectors in a finite dimensional
Hilbert space. This is in the same spirit as the ``quantum nets''
introduced in \cite{GHW:2003}, where a map between the blocks of a
design to vectors in Hilbert space with prescribed inner products was
used to define a discrete Wigner transform (see also
\cite{Galvao:2004,PRS:2004,PR:2005,Paz:2002}).

Let $(X,\mathcal{B})$ be an affine $(v,b,r,k,\lambda)$ design.  We
associate with each block $B$ in $\mathcal{B}$ a vector of unit norm
in $\C^v$, and we denote this vector by $\ket{B}$.  We encode the
information about the parallel classes of the design in the angles
between the vectors. We require from the vectors $\ket{B}$ with $B\in
\mathcal{B}$ that they satisfy 
\begin{equation}\label{eq:angle}
\langle B | C \rangle = \left\{
\begin{array}{rl} 
\delta_{B,C} & \mbox{if $B$ and $C$ are
parallel}, \\
k/v & \mbox{if $B$ and $C$ are not parallel}. 
\end{array}\right.
\end{equation}
We will call a vector system $\{ \ket{B}\,|\, B\in \mathcal{B}\}$ that
satisfies the constraints (\ref{eq:angle}) a \textsl{realization} of the
affine design $(X,\mathcal{B})$.

\begin{lemma}
An affine $(v,k,\lambda)$ design $(X,\mathcal{B})$ has a realization
in~$\C^v$.
\end{lemma}
\begin{proof}
Suppose that $v_B$ is the incidence vector of the block $B$.  Then
$\ket{B}=\frac{1}{\sqrt{k}}v_B$, with $B\in \mathcal{B}$, is a realization of $X$ in
$\C^v$. \end{proof}

\begin{example}
Let $n$ be a power of a prime, and let $(X,\mathcal{B})$ be an affine
plane of order $n$. Suppose that $\mathcal{B}$ is the disjoint union
of the parallel classes $L_m$, $\mathcal{B}=L_0\cup\cdots \cup L_n$.
Let $B_a=\{v_{a,b} \,|\, 1\le b\le n\}$, with $0\le a\le n$,
denote a set of $n+1$ mutually unbiased bases of $\C^n$.  Suppose that
the parallel class $L_a$ is given by the set of lines $L_a=\{
\ell_{a,b}\,|\, 1\le b\le n\}$. If we define 
\begin{equation}\label{eq:mub} 
\ket{\ell_{a,b}} = v_{a,b}\otimes \overline{v_{a,b}}\in \C^{n^2},
\end{equation} then 
$\bracket{\ell_{a,b}}{\ell_{a',b'}}=1/n=n/n^2$ if the lines are from
different parallel classes, $L_a\neq L_{a'}$; and
$\bracket{\ell_{a,b}}{\ell_{a,b'}}=\delta_{b,b'}$ for parallel lines.
The vectors in~(\ref{eq:mub}) are a realization of the affine plane in
$\C^{n^2}$.
\end{example}

\myparagraph{Reconstruction.}  We saw in the previous two sections how
one can explore properties of the previous example to solve the first
two problems of the Mean King.  A far reaching generalization is
provided by the realizations of affine designs. We begin with a
generalization of Corollary~\ref{basis}.

\begin{theorem}\label{statesTheorem}
Let $(X, \mathcal{B})$ be an affine $(v,b,r,k,\lambda)$ design.
Suppose that ${\cal C}$ is an arbitrary parallel class of $X$, and $\{
\ket{B} : B \in \mathcal{B}\}$ a realization of $X$ in~$\C^v$.
If we define 
\begin{equation}\label{states}
\ket{\psi_p} = - \alpha \left(\sum_{B \in {\cal C}} \ket{B}\right) 
+ \beta \left(\sum_{B : p \in B} \ket{B} \right),
\end{equation}
with $\alpha = (r-1){\sqrt{k}}/{v}$ and $\beta =1/\sqrt{k}$, then $\{ \ket{\psi_p} : p \in X\}$ forms an
orthonormal basis of $\C^v$.
\end{theorem}
\begin{proof}
Let $p$, $q$ be two arbitrary points in $X$. Before we
compute the inner product of the states $\ket{\psi_p}$ and
$\ket{\psi_q}$ we first consider the following intersection
scenarios. 

Let $P=\{B\in \mathcal{B}\,|\, p\in B\}$ and $Q=\{C\in
\mathcal{B}\,|\, q\in C\}$. We shall be interested in the inner
products $\langle B | C \rangle$ with $B \in P$ and $C \in Q$.  Since
each point is incident with $r$ blocks, we obtain precisely $r^2$ such
pairs. In the case $p\neq q$, we get 
\begin{compactenum}[a)]
\item $\bracket{B}{C}=k/n$ for $r(r-1)$ pairs $(B,C)$, since there are
$r(r-1)$ ways to choose pairs $(B,C)$ such that $B$ and $C$ belong to
different parallel classes;
\item $\bracket{B}{B}=1$ for $\lambda$ pairs $(B,B)=(B,C)$, since there 
are $\lambda$ blocks $B$ in the intersection $P\cap Q$; 
\item $\bracket{B}{C}=0$ for the remaining $r-\lambda$ pairs $(B,C)$ 
that are in the same parallel class, but satisfy $B\cap C=\emptyset$. 
\end{compactenum}
In the case $p=q$, we obtain
\begin{compactenum}[a)]
\item $\bracket{B}{B}=1$ for $r$ pairs of the form $(B,B)=(B,C)$, 
where $B$ and $C$ belong to the same parallel class. 
\item $\bracket{B}{C}=k/n$ for the remaining $r(r-1)$ pairs, where $B$ and $C$ belong to different parallel classes. 
\end{compactenum}
Suppose that we have a point $p$ and a parallel
class $\mathcal{C}$. We are interested in the values of the 
inner products $\langle
B | C \rangle$ of pairs $(B,C)$ with $B \in P$ and $C \in {\cal
C}$. Since the parallel class $\mathcal{C}$ contains $v/k$ blocks, we
obtain $rv/k$ such pairs, and 
\begin{compactenum}[a)]
\item $\bracket{B}{C}=1$ for one pair $(B,B)=(B,C)$ with $B\in \mathcal{C}$; 
\item $\bracket{B}{C}=0$ for the $v/k-1$ pairs $(B,C)$  with $B\in \mathcal{C}$ and $B\cap C=\emptyset$; 
\item $\bracket{B}{C}=k/n$ for the remaining $(r-1)v/k$ pairs with $B\not\in \mathcal{C}$. 
\end{compactenum}
\smallskip

These design-theoretic arguments enable us to compute the 
inner product $\bracket{\psi_p}{\psi_q}$ of two states, 
\small 
\begin{eqnarray*}
\langle \psi_p | \psi_q \rangle &=& \alpha^2 \sum_{B,B^\prime \in {\cal C}}
 \langle B | B^\prime \rangle - 2 \alpha \beta \sum_{B \in {\cal C}}
\sum_{B^\prime : p \in B^\prime}   \langle B | B^\prime \rangle + 
\beta^2 \sum_{B : p \in B}
\sum_{B^\prime : q \in B^\prime}   \langle B | B^\prime \rangle \\
& = & 
\alpha^2 \left(\frac{v}{k}\right) - 2 \alpha \beta \left(1 + 
\left(\frac{v}{k}-1\right)\cdot 0 + \frac{v}{k} (r-1)
 \frac{k}{v}\right) \\
& &  \qquad + 
\left\{ \begin{array}{ccc}
\beta^2 \left(\lambda + (r-\lambda) \cdot 0 + (r^2-r) \frac{k}{v} \right)
& \text{if } p \not= q, \\[1ex]
\beta^2 \left(r + (r^2-r) \frac{k}{v} \right) & \text{if } p = q. 
\end{array}\right.
\end{eqnarray*}\normalsize 
We now solve this for $\alpha$ and $\beta$ with respect to the
constraints $\langle \psi_p | \psi_q \rangle = \delta_{p,q}$ and show that
the given values for $\alpha$ and $\beta$ indeed are solutions. 

Subtracting the first equation from the second yields
$\beta^2 ( r - \lambda) = 1$. We know from Lemma~\ref{parameters} (iii) that  
$k=r-\lambda$; hence, $\beta = 1/\sqrt{k}$. 
Therefore, the quadratic equation for $\alpha$ simplifies to  
\[
\left(\frac{v}{k}\right) \alpha^2 - \left(\frac{2r}{\sqrt{k}}\right) \alpha + \left(
\frac{\lambda}{k} + \frac{r^2-r}{v} \right) = 0.
\]
Solving for $\alpha$, we obtain 
\begin{eqnarray*}
\alpha_{1,2} &=& \left(\frac{2r}{\sqrt{k}} \pm 
\sqrt{ \frac{4r^2}{k} - 4 \frac{v}{k} \left( \frac{\lambda}{k} +
\frac{r^2-r}{v} \right)}\right)\!\bigg/\!\left(\frac{2v}{k}\right) \\
&=& \left(\frac{2r}{\sqrt{k}} \pm 
\sqrt{ - 4\frac{v \lambda}{k^2} + 4 \frac{r}{k}}\right)\!\bigg/\!
\left(\frac{2v}{k}\right) 
\end{eqnarray*}
By Lemma~\ref{parameters} (ii), we have $v\lambda =k\lambda +
k(k-1)$. If we subsitute this relation into the previous expression,
then we can simplify the solution further by taking the relation 
$k=r-\lambda$ into account:
\begin{eqnarray*}
\alpha_{1,2}&=& \left(\frac{2r}{\sqrt{k}} \pm 
\sqrt{ - \frac{4(\lambda k + k (k-1))}{k^2} + 4 \frac{r}{k}}\right)\!\bigg/\!
\left(\frac{2v}{k}\right) \\
&=& \left(\frac{2r}{\sqrt{k}} \pm 
\sqrt{ \frac{4}{k}(r-\lambda) - 4 \frac{k-1}{k}}\right)\!\bigg/\!
\left(\frac{2v}{k}\right) \\
& = & \frac{2}{\sqrt{k}} (r\pm 1) \frac{k}{2v} = \frac{\sqrt{k}}{v}(r
\pm 1).
\end{eqnarray*}
Since we have chosen $\alpha = (r-1)\sqrt{k}/v$ and $\beta
=1/\sqrt{k}$, it follows from our calculation that
$\{\ket{\psi_p}\,|\,p \in X\}$ is an orthonormal basis of~$\C^v$.
\end{proof}
\smallskip

A remarkable property of the basis given in the previous theorem is
that if we measure a state $\ket{B}$, with $B\in \mathcal{B}$, then we
will only observe points $p$ that are incident with $B$, that is, the
state $\ket{B}$ can only collapse to $\ket{\psi_p}$ with $p\in B$.
The reader should contrast the next theorem with Lemma~\ref{extraction},
which serves the same purpose in the case of affine planes.

\begin{theorem}\label{th:adesign}
Let $(X, \mathcal{B})$ be an affine $(v,b,r,k,\lambda)$
design. Suppose that $\{ \ket{B} : B \in \mathcal{B}\}$ is a
realization of $X$ in $\C^v$, and let $\{ \ket{\psi_p} : p \in X\}$ be
the associated orthonormal basis given in Theorem \ref{statesTheorem}.
If\/ $B \in {\cal B}$, then $\langle \psi_p | B \rangle \not = 0$ if
and only if $p \in B$.
\end{theorem}
\begin{proof}
Let ${\cal C}$ be an arbitrary parallel class of
$X$. Note that the normalized state $\ket{\varphi_{\cal C}} = 1/\sqrt{|{\cal C}|} \sum_{B \in {\cal C}} \ket{B}$ has the
property that $\ket{\varphi_{\cal C}} = \ket{\varphi_{{\cal C}^\prime}}$ for
any other parallel class ${\cal C}^\prime$. Indeed, the computation of
the inner product of two such states reveals that 
\begin{eqnarray*}
\langle \varphi_{\cal C} | \varphi_{{\cal C}^\prime}
\rangle &=& \frac{1}{|C|} \sum_{B \in {\cal C}} \sum_{B^\prime 
\in {\cal C}^\prime} \langle B | B^\prime \rangle \\
& = &\frac{1}{|C|} \cdot |C|^2 \cdot \frac{k}{v} = 1.
\end{eqnarray*}
Now, let $B_0 \in \mathcal{B}$ be an arbitrary block. We distinguish the
two cases (i)  $p \in B_0$ and (ii) $p \notin B_0$. First, in case (i) we
assume that $p\in B_0$ and let ${\cal C}$ be the parallel class
containing $B$. We obtain that
\begin{eqnarray*}
\langle \psi_p | B_0 \rangle &=&
-\alpha \sum_{B \in {\cal C}} \langle B | B_0 \rangle
+ \beta \sum_{B:  p \in B} \langle B | B_0 \rangle \\
& = & - \alpha + \beta \cdot r \cdot \frac{k}{v}  = -
\frac{\sqrt{k}}{v}(r-1) + \frac{rk}{\sqrt{k}v} = \frac{\sqrt{k}}{v}
\not=0.
\end{eqnarray*}
In case (ii) we have that $p \notin B_0$. We now apply Lemma
\ref{blockPoint} and obtain that there is precisely one block which
contains $p$ and which is disjoint from $B_0$. Of all the $r$ blocks
which pass through $p$ all the other $r-1$ ones intersect
nontrivially with $B_0$. Hence, we obtain that in this case
\begin{eqnarray*}
\langle \psi_p | B_0 \rangle &=&
-\alpha \sum_{B \in {\cal C}} \langle B | B_0 \rangle
+ \beta \sum_{B:  p \in B} \langle B | B_0 \rangle \\
& = & - \alpha + \beta \cdot (r-1) \cdot \frac{k}{v}  -\frac{\sqrt{k}}{v}(r-1) + \frac{\sqrt{k}}{v}(r-1) = 0.
\end{eqnarray*}
Hence, a measurement of $\ket{B_0}$ in the basis given by the vectors
$\ket{\psi_p}$ can only yield a result for values of $p$ such that $p
\in B_0$. 
\end{proof}

\myparagraph{Generic King's Problem.}  The previous two theorem are
the key to a much more general class of Mean King's problems. We
briefly sketch the idea of the generic version, and then illustrate it
with an example in the next section.  Suppose that $(X,\mathcal{B})$
is an affine $(v,b,r,k,\lambda)$ design, and the vector system
$\left\{\, \ket{B}\,|\, B\in \mathcal{B}\right\}$ is a realization of $X$ in $\C^v$.

\begin{enumerate}
\item Alice constructs the state
$\varphi=\frac{1}{\sqrt{|C|}}\sum_{B\in \mathcal{C}} \ket{B}$ for some
parallel class $\mathcal{C}$.
\item Alice hands the state to the King. 
\item The King's men perform a measurement on a subsystem that
corresponds to one of the parallel classes $\mathcal{C}'$ of $X$, so
that the state collapses to $\ket{B}$ with $B\in \mathcal{C}'$. 
\item The King's men hand the quantum system back to Alice. And Alice
performs a von Neumann measurement with respect to the basis
$\ket{\psi_p}$, $p\in X$.  She will only observe points $p$ that are
incident with the block $B$.
\item The King reveals the measurement or, equivalently, the parallel
class $\mathcal{C}'$. Alice simply checks which block $B'$ in
$\mathcal{C}'$ contains the point $p$, and announces that block $B'$.
The block $B'$ derived by Alice and the block $B$ observed by the
King's men must coincide, because precisely one block of a parallel
class contains $p$.
\end{enumerate}
Step 3 is quite ambiguous and the designer of problem has considerable
freedom to realize this requirement.

\mysection{Solution to the Third Problem}\nopagebreak
\begin{tale} Since Alice was: very familiar with the work of Jacques: 
Hadamard, s:he quickly had an idea as: to which quantum s:tate s:he might
prepare. This: time s:he chos:e an entangled s:tate of the three
atoms: which has: the property that no matter which of the nine
meas:urements: the King performs:, the res:ults: can be
dis:tinguis:hed from her own meas:urement data and the King's:
revelation of the bas:is:. She pas:s:ed the tes:t with flying colours:
and left the is:land and the flabbergas:ted King behind.
\end{tale}

\myparagraph{Hadamard Designs.}  We denote the transpose of a matrix
$A$ by $A^t$.  Recall that a Hadamard matrix or order $n$ is a $\pm 1$
matrix $H_n$ of size $n\times n$ with the property that $H_n H_n^t = n
\onemat_n$. A necessary condition for the existence of a Hadamard
matrix is that either $n=2$ or $n\equiv 0 \; {\rm mod}\; 4$. A
long-standing conjecture is that Hadamard matrices exist for all such
$n$, but a proof is elusive; see~\cite{BJL:99I,hedayat99,Stinson:2003} 
for a wealth of constructions of Hadamard matrices. 

The Hadamard matrix $H_2= \tiny\left(\begin{array}{@{}r@{}r@{}} 1 & 1
\\ 1 & -1\end{array}\right)$, and the tensor product of Hadamard
matrices is again a Hadamard matrix; hence, there exist Hadamard
matrices $H_{2^k}$ for $k\ge 1$. In particular, a Hadamard matrix for
$n=8$ is given by 
{\small 
\begin{equation}\label{had8}
{\normalsize H_8} = \left(
\begin{array}{cccccccc} 
+ & + & + & + & + & + & + & + \\ 
+ & - & + & - & + & - & + & - \\ 
+ & + & - & -
& + & + & - & - \\ + & - & - & + & + & - & - & + \\ + & + & + & + & -
& - & - & - \\ + & - & + & - & - & + & - & + \\ + & + & - & - & - & -
& + & + \\ + & - & - & + & - & + & + & -
\end{array}
\right).
\end{equation}}%
where the entries $\pm 1$ have been abbreviated to $+/-$. 
We obtain a design as follows. Define the set of points to be 
$X=\{1,\dots, 8\}$. The blocks are obtained from the rows of $H_8$ which are
different from the all-ones row. For each row we obtain two blocks by
grouping the elements which are respectively labeled ``$+$'' and ``$-$''
together. Explicitly, we obtain the blocks
\[
\begin{array}{c@{\;,\quad}c}
B_{1}^{+} = \{1, 3, 5, 7\} & B_{1}^- = \{2, 4, 6, 8\}, \\
B_{2}^{+} = \{1, 2, 5, 6\} & B_{2}^- = \{3, 4, 7, 8\}, \\
B_{3}^{+} = \{1, 4, 5, 8\} & B_{3}^- = \{2, 3, 6, 7\}, \\
B_{4}^{+} = \{1, 2, 3, 4\} & B_{4}^- = \{5, 6, 7, 8\}, \\
B_{5}^{+} = \{1, 3, 6, 8\} & B_{5}^- = \{2, 4, 5, 7\}, \\
B_{6}^{+} = \{1, 2, 7, 8\} & B_{6}^- = \{3, 4, 5, 6\}, \\
B_{7}^{+} = \{1, 4, 6, 7\} & B_{7}^- = \{2, 3, 5, 8\}. \\
\end{array}
\]
The blocks $B_{i}^{+}$ and $B_{i}^{-}$ are parallel for $1\le i\le 7$,
and blocks $B_i^\pm$ and $B_k^\pm$ with $i\neq j$ intersect in 2
points. Put differently, we have obtained an affine 
$(8,14,7,4,3)$ design. 
\smallskip

\textit{Remark.}  If there exists a Hadamard matrix $H_n$ of size $n$,
then there exists an affine $(n,2n-2,n-1,n/2, n/2-1)$ design,
see~\cite[p.~110]{Stinson:2003}.
\smallskip

\nix{
Indeed,
assume without loss of generality that the first row $r_0$ of $H_n$ is
the all-ones row.  For each row $r_i$ of $H_n$, with $1\le i< n$ of
$H_n$, we obtain two blocks $B_i^+$ and $B_i^-$, namely the collection
of points $j$ for which respectively $r_i(j) = 1$ and $r_i(j) -1$. Clearly, this gives a total of $n-1$ points and $2n-2$
blocks. Two blocks are parallel if and only if they belong to the same
row of $H_n$. Since each point corresponds to a unique column of $H_n$,
we obtain that each point is incident with $n-1$ blocks. Each block
has size $k=n/2$, since the rows have to be orthogonal to the all-ones
vector. Finally, the intersection of any two non-parallel blocks
contains $\lambda=n/4$ elements, since two rows of $H_n$ have to be
orthogonal.
}

\myparagraph{Representation.}  Let us define a realization of the
affine $(8,14,7,4,3)$ design $X$ in~$\C^8$ by the vectors 
\newcommand{\ppi}{\phantom{i}}
$$\small  
\begin{array}{l}
\begin{array}{l@{}c@{}l@{,\qquad\,\!\;\qquad}l@{}c@{}l@{,}}
\ket{B^+_1} &=& \frac{1}{\sqrt{2}}(\ket{000} + \ket{011}) & 
\ket{B^-_1} &=& \frac{1}{\sqrt{2}}(\ket{101} + \ket{110}) \\[2ex]
\ket{B^+_2} &=& \frac{1}{\sqrt{2}}(\ket{000} + \ket{101}) &
\ket{B^-_2} &=& \frac{1}{\sqrt{2}}(\ket{011} + \ket{110})  \\[2ex]
\ket{B^+_3}&=& \frac{1}{\sqrt{2}}(\ket{000} + \ket{110}) & 
\ket{B^-_3}&=& \frac{1}{\sqrt{2}}(\ket{011} + \ket{101}) \\[2ex]
\end{array}\\[5mm]
\begin{array}{l@{}c@{}l}
\ket{B^+_4} &=& \frac{1}{2\sqrt{2}}(\ket{000} + \ppi \ket{001}
+ \ppi \ket{010} +\ppi \ket{011} + \ppi \ket{100} + \ppi \ket{101} +
\ppi \ket{110} + \ppi \ket{111}),\\[2ex]
\ket{B^-_4} &=& \frac{1}{2\sqrt{2}}(\ket{000} - \ppi \ket{001}
- \ppi \ket{010} + \ppi \ket{011} - \ppi \ket{100} + \ppi \ket{101} +
\ppi \ket{110} - \ppi \ket{111}),\\[2ex]
\ket{B^+_5} &=& \frac{1}{2\sqrt{2}}(\ket{000} -i \ket{001} -i \ket{010} 
+ \ppi \ket{011} +i \ket{100} + \ppi \ket{101} + \ppi \ket{110} + 
i \ket{111}),\\[2ex]
\ket{B^-_5} &=& \frac{1}{2\sqrt{2}}(\ket{000} + i \ket{001} +i \ket{010} 
+ \ppi \ket{011} -i \ket{100} + \ppi \ket{101} + \ppi \ket{110} 
-i \ket{111}),\\[2ex]
\ket{B^+_6} &=& \frac{1}{2\sqrt{2}}(\ket{000} -i \ket{001} +i \ket{010} 
+ \ppi \ket{011} -i \ket{100} + \ppi \ket{101} + \ppi \ket{110} 
+i \ket{111}),\\[2ex]
\ket{B^-_6} &=& \frac{1}{2\sqrt{2}}(\ket{000} +i \ket{001} -i \ket{010} 
+ \ppi \ket{011} +i \ket{100} + \ppi \ket{101} + \ppi \ket{110} 
-i \ket{111}),\\[2ex]
\ket{B^+_7} &=& \frac{1}{2\sqrt{2}}(\ket{000} + i \ket{001} - i \ket{010} 
+ \ppi \ket{011} - i \ket{100} + \ppi \ket{101} + \ppi \ket{110} 
+ i \ket{111}),\\[2ex]
\ket{B^-_7} &=& \frac{1}{2\sqrt{2}}(\ket{000} - i \ket{001} + i \ket{010} 
+ \ppi \ket{011} + i \ket{100} + \ppi \ket{101} + \ppi \ket{110} 
- i \ket{111}).
\end{array}
\end{array}
$$ 
Recall that the parallel classes $\mathcal{C}_k$ of the design $X$ 
are given by $\mathcal{C}_k=\{ B^+_k,B^-_k\}$ for $1\le k\le 7$.
One can check that 
{\small $$
\langle B^+_i | B^-_i \rangle = 0, \quad
\langle B^+_i | B^-_j \rangle = 1/2,\quad
\langle B^+_i | B^+_j \rangle = 1/2, \quad
\langle B^-_i | B^-_j \rangle = 1/2,
$$}%
holds for distinct $i$ and $j$ in the range $1\le i,j\le 7$, so the
system of vectors forms indeed a realization of the affine $(8,14,7,4,3)$
design~$X$.

\myparagraph{The King's Measurements.} Alice prepares the state 
\begin{equation}\label{GHZtransformed}
\ket{\varphi} = \frac{1}{\sqrt{2}}(\ket{B^+_k}+\ket{B^-_k} 
= \frac{1}{2}(\ket{000} +
\ket{011} + \ket{101} + \ket{110}).
\end{equation}
We note that $\ket{\varphi}$ can be obtained from the 
GHZ state
$\frac{1}{\sqrt{2}}(\ket{000} + \ket{111})$
by applying a Hadamard gate to each qubit~\cite{GHZ:89}.

The King can perform nine different measurements; namely, he can
measure one of the three spin components of either the first, second,
or third qubit. The three measurements are performed with respect to
standard basis $B_s$, the Hadamard basis $B_h$ or a third
complementary basis $B_u$; explicitly,  
$$ 
\begin{array}{c}
B_s=\textstyle\big\{\ket{0},\ket{1}\big\},\quad 
B_h=\big\{
\frac{1}{\sqrt{2}}\ket{0}+\frac{1}{\sqrt{2}}\ket{1}, 
\frac{1}{\sqrt{2}}\ket{0}-\frac{1}{\sqrt{2}}\ket{1}\big\},\\[2ex]
B_u=\big\{
\frac{1}{\sqrt{2}}\ket{0}+\frac{i}{\sqrt{2}}\ket{1}, 
\frac{1}{\sqrt{2}}\ket{0}-\frac{i}{\sqrt{2}}\ket{1}\big\}.
\end{array}
$$ 
The corresponding projectors on one qubit are respectively given by 
\newcommand{\phh}{\phantom{\frac{1}{2}}}
\[
\begin{array}{l@{\,}c@{\,}l@{,\quad}l@{\,}c@{\,}l}
P^+_s &=& \phh\left(\begin{array}{rr@{}} 1 & 0 \\ 0 & 0 \end{array}\right) &
P^+_s &=& \phh \left(\begin{array}{@{}rr@{}} 0 & 0 \\ \phantom{-}0 & 1 \end{array}\right), \\[1ex]
P^+_h &=& \frac{1}{2} \left(\begin{array}{rr@{\,}}  1 &  1 \\ 1 & 1 
\end{array}\right) &
P^-_h &=& \frac{1}{2} \left(\begin{array}{@{}r@{\,}r@{}}  1 & -1 \\ -1 & 1 
\end{array}\right), \\[1ex] 
P^+_u &=& \frac{1}{2} \left(\begin{array}{@{}rr@{}}  1 & -i \\ i & 1 
\end{array}\right) &
P^-_u &=& \frac{1}{2} \left(\begin{array}{@{}rr@{}} 1 & i \\ -i & 1 
\end{array}\right).
\end{array}
\]
It is straightforward to verify that the nine measurements corrspond
the seven parallel classes $\mathcal{C}_1,\dots, \mathcal{C}_7$ as
follows: A measurement in the standard basis on qubit $k$ amounts to
applying $P^+_s$ or $P^-_s$ to any of the three qubits and collapses
$\ket{\varphi}$ to either $\ket{B^+_k}$ or $\ket{B^-_k}$, where $k \in
\{1, 2, 3\}$. The three Hadamard measurements amount to applying
$P^+_h$ or $P^-_h$ and map the state $\ket{\varphi}$ either to
$\ket{B^+_4}$ or to $\ket{B^-_4}$. Finally, the three measurements in
the basis $B_u$ amount to applying $P^+_u$ or $P^-_u$ and map the state
$\ket{\varphi}$ to either $\ket{B^+_k}$ or $\ket{B^-_k}$, where $k \in
\{5, 6, 7\}$.  Table~\ref{hadamard} summarizes the King's measurement
outcomes.

\begin{table}[t]
\begin{center}
\small\begin{tabular}{|c||c|c|}
\hline
Measurement & Outcome ``0'' & Outcome ``1'' \\
\hline
\hline
$M_{1,s}$ & $\ket{B^+_1}$ & 
$\ket{B^-_1}$ \\ \hline
$M_{2,s}$ & 
$\ket{B^+_2}$ & 
$\ket{B^-_2}$ \\ \hline
$M_{3,s}$& 
$\ket{B^+_3}$& 
$\ket{B^-_3}$  \\ \hline
$M_{1,h}$, $M_{2,h}$, $M_{3,h}$ &  $\ket{B^+_4}$ 
& $\ket{B^-_4}$ \\ \hline
$M_{1,u}$& $\ket{B^+_5}$ & 
$\ket{B^-_5}$ \\ \hline
$M_{2,u}$& 
$\ket{B^+_6}$ & 
$\ket{B^-_6}$ \\ \hline
$M_{3,u}$& 
$\ket{B^+_7}$& 
$\ket{B^-_7}$  \\
\hline
\end{tabular}
\caption{The table shows the resulting states after measuring
$\ket{\varphi}$ in one of the nine different bases.  We use $M_{k,s}$
to denote that the $k$th qubit has been measured with respect to the
standard basis $B_s$, we use $M_{k,h}$ to denote that
the $k$th qubit has been measured with respect to the Hadamard basis
$B_h$, and we denote $M_{k,u}$ to denote that the
$k$th qubit has been measured with respect to the third complementary
basis $B_u$. Note that the resulting states for $M_{1,h}$, $M_{2,h}$,
and $M_{3,h}$ are the same.  }
\label{hadamard}
\end{center}
\end{table}

\myparagraph{Alice's Measurement.}  The key to Alice's success is that
she devices a measurement that allows her to infer a point $p$ that is
incident with the block that the King's men have measured. If the King
later reveals the measurement, or parallel class, then Alice simply
needs to identify the block that contains $p$.

Alice can use Theorem~\ref{statesTheorem} to construct a suitable
orthonormal basis. Since the parameters of our design are $v=8$,
$r=7$, and $k=4$, we find that $\alpha = (r-1)\sqrt{k}/{v} = 3/2$ and
$\beta = 1/\sqrt{k} = 1/2$.  We can compute the states
$$\ket{\psi_p} = - \frac{3}{2} \left(\sum_{B \in {\cal C}} \ket{B}\right) +
\frac{1}{2} \left(\sum_{B : p \in B} \ket{B} \right).$$  Explicitly, we get 
\[ 
\ket{\psi_{1,5}} = \frac{1}{\sqrt{2}}(\ket{000} \pm \ket{001}), \quad
\ket{\psi_{2,6}} = \frac{1}{\sqrt{2}}(\ket{110} \pm \ket{100}),
\]
\[
\ket{\psi_{3,7}} = \frac{1}{\sqrt{2}}(\ket{011} \pm \ket{010}), \quad
\ket{\psi_{4,8}} = \frac{1}{\sqrt{2}}(\ket{101} \pm \ket{111}).
\]
The result obtained from this measurement corresponds to a point $p
\in \{1, \ldots, 8\}$. If now the King reveals his measurement,
i.\,e., discloses which qubit was measured and if the standard basis
or Hadamard basis was used, then this uniquely specifies a parallel
class ${\cal C}_1, \ldots, {\cal C}_7$. Now, the correct outcome ``0''
or ``1'' of the King's measurement is given by the unique block $B$ in
this class, such that $p\in B$.

For instance, starting from $\ket{\varphi}$, assume that the King
decides the measure the second qubit in the Hadamard basis.  Assume
that his measurement result was ``1''. Then the state has collapsed to
$\ket{B^-_4}$, where $B^-_4=\{5,6,7,8\}$.  Alice will now measure one
of $\ket{\psi_5}$, $\ket{\psi_6}$, $\ket{\psi_7}$, $\ket{\psi_8}$, the
other states do not occur in her measurement. Now, if the King
discloses that he performed a measurement of the second qubit in the
Hadamard basis, then Alice knows that this measurement yields either
$\ket{B^+_4}$ or $\ket{B^-_4}$. The point $p$ that Alice had measured
is an element of $\{5,6,7,8\}$, and either choice yield the block
$B^-_4$, the block corresponding to the result ``1''.

\paragraph{Acknowledgments.}
The research of A.K. was supported by NSF CAREER award CCF-0347310,
NSF grant CCR-0218582, a TEES Select Young Faculty award, and a Texas
A\&M TITF grant. This work was carried out while M.R. was at the
Institute for Quantum Computing, University of Waterloo, Canada.

We thank Thomas Beth, Mauro D'Ariano, Ernesto Galvao, Chris Godsil,
Markus Grassl, and Aidan Roy several interesting discussions.  A very
special thanks to Yannis Haralambous for designing such beautiful
fonts! 
\smallskip

We apologize for taking the typographical liberty to typeset
{\swabfamily passable} as {\swabfamily pas:s:able}, but we found that
most readers have difficulties to {\swabfamily
distinguish} the former version from {\swabfamily paffable}.

\normalsize 

\end{empfile}

\noindent $^*$Department of Computer Science, Texas A\&M 
University, College Station, TX 77843-3112, USA,
\texttt{klappi <at-sign> cs.tamu.edu}\\

\noindent $^\dagger$NEC Laboratories America, Inc., 
4 Independence Way, Princeton, NJ 08540, USA, 
\texttt{mroetteler <at-sign> nec-labs.com}\\

\begin{quote}
\noindent\textbf{Summary.} The Mean King's problem asks to determine
the outcome of a measurement that is randomly selected from a set of
complementary observables. We review this problem and offer a
combinatorial solution. More generally, we show that whenever an
affine resolvable design exists, then a state reconstruction problem
similar to the Mean King's problem can be defined and solved. As an
application of this general framework we consider a problem involving
three qubits in which the outcome of nine different measurements can
be determined without using ancillary qubits.  The solution is based
on a measurement derived from Hadamard designs.
\end{quote}

\end{document}